# Unsupervised Neural Network-Naive Bayes Model for Grouping Data Regional Development Results

Azhari SN
Computer Science and Electronic Dept.
Universitas Gajah Mada
Yogyakarta - Indonesia

Tb. Ai Munandar
Eng. Informatics Dept.
Universitas Serang Raya
Serang – Banten - Indonesia

## ABSTRACT
Determination quadrant development has an important role in order to determine the achievement of the development of a district, in terms of the sector's gross regional domestic product (GDP). The process of determining the quadrant development typically uses Klassen rules based on its sector GDP. This study aims to provide a new approach in the conduct of regional development quadrant clustering using cluster techniques. Clustering is performed based on the average value of the growth and development of a district contribution compared with the average value and contribution of the development of the province based on data in comparison with a year of data to be compared. Testing models of clustering, performed on a dataset of two provinces, namely Banten (as a data testing) and Central Java (as the training data), to see the accuracy of the classification model proposed. The proposed model consists of two learning methods in it, namely unsupervised (Self Organizing Map / SOM-NN) method and supervised (Naive Bayess). SOM-NN method is used as a learning engine to generate training data for the target Class that will be used in the machine learning Naive Bayess. The results showed the clustering accuracy rate of the model was 98.1%, while the clustering accuracy rate of the model results compared to manual analysis shows the accuracy of the typology Klassen smaller, ie 29.63%. On one side, clustering results of the proposed model is influenced by the number and keagaraman data sets used

## 1. INTRODUCTION
Development planning involves essentially seek and seek harmony and balance between regions according to its potential, so that they can be utilized in an efficient, safe and orderly [1]. In addition, the development should have a clear direction of equity and sustainable to meet the needs of people with fair and equitable distribution [2].

Equitable development would be key to the success of economic development in Indonesia so that each region is expected to have a faster development in accordance with its powers [3]. One way to measure equitable development is the area of the quadrant grouping the construction area to see the extent to which the achievement of the development of the area in question using the typology Klassen. Typology Klassen is a common tool used to analyze the state of the regional economy of an area resulting classification is based on the construction of sub-sector regional gross domestic product - GDP [4]. Klassen analysis, classifying the data sub-sector GDP into four groups of level of development of a region, based on statistical data on nine components of GDP, namely Quadrant I, an advanced sector and rapid growth, Quadrant II, advanced but depressed sector, Quadrant III, sector potential or, still can develop and Quadrant IV, is relatively underdeveloped sector [4],[5].

On the other hand, the development of information technology has made many contributions, especially in the process of grouping the data, so it can be used to help classify data from the development of an area, so it can be used to determine the level of development of the area concerned. Many methods can be used to process the data grouping, either supervised or unsupervised.

This study discusses the process of grouping the data from the results of the development of GDP by sector so that it can be used to determine the level of development of the region. The approach used unsupervised and supervised methods for grouping data. Unsupervised clustering method used is self-organizing maps neural network (SOM-NN) to form a development class in which the result of the formation of SOM-NN Class, Class used as a target in the supervised method, namely Naive bayess. Both methods are combined to form a classification model that is capable of grouping the data on the development of an area.

Some of previous studies using the concept of SOM for many grouping needs, for example for the grouping of digital image processing, both for segmentation [6],[7], compress the image without changing the quality of the input image while still producing good quality image compressed [8], clustering of documents [9],[10] the determination of a microbial taxonomy relation class [11], defining a strategy for grouping customer market share [12], analysis of strategic groups of construction companies to understand the strategic position of the company [13], predictive classification of computer network attacks [14], visualization of spatial data to find structure and pattern of data in order to obtain new information relationships between socio-economic indicator data of an area [15], the grouping capital road users based on peak and off peak time so it can be used as decision support in planning the construction of transportation facilities [16] can even be used to predict the possible locations of clarifying bedasarkan aftershock earthquake data trends in a region [17].

Similarly, the SOM-NN, Naive Bayes concept is widely used for classification needs in various fields, such as web pages classification training [18], classification of data for disease diagnosis needs [19], text classification [20] and documents [21].

This study is divided into five sections, the first section discusses the research background, the second section discusses the literature review and theory used in this study, the third part is the research methodology used and explains the research flow, the fourth section is a discussion of the results of research that contains a description of proposed model and simulation testing the model against the data sector GDP, and the last is the cover that contains conclusions and suggestions for further research.

## 2. KLASSEN TIPOLOGY

Klassen typology is an analytical tool that can be used to see patterns and economic development of a region, seen from forming sector Gross Regional Domestic Product (GDP), which would classify a region into four major quadrants, ie areas with rapid growth, the depressed growth, areas that can still be developed and relatively underdeveloped regions [4],[5]. Here is a table Klassen typology classification (see table 1):

**Table 1. Classification Matrix of Economic Growth According to Klassen Typology**

| Kuadran I<br>an advanced sector and rapid growth (developed sector)<br>$r_i > r$ dan $y_i < y$ | Kuadran II<br>advanced but depressed sector (stagnant sector)<br>$r_i < r$ dan $y_i > y$ |
|---|---|
| Kuadran III<br>Potential sector<br>(developing sector)<br>$r_i > r$ dan $y_i > y$ | Kuadran IV<br>relatively underdeveloped sector<br>$r_i < r$ dan $y_i < y$ |

Remarks:
$r_i$ = rate of growth of GDP District *i*
$r$ = rate of growth of total GRDP
$y_i$ = income per capita district *i*
$y$ = income per capita provincial

## 3. NAIVE BAYESS

Naive Bayes classifier is a method that is based on the assumption that the probability of each variable X is independence. This method assumes that the existence of each attribute (variable) has nothing to do with the existence of attributes (variables) in the other.

Naive Bayes is the basis of the Bayes theorem, which states that, if X is a class of data samples (a label) is not known, and H is the hypothesis where X is a class of data (labels) C, and P (H) is the chance of a hypothesis H, then P (X) is expressed as a chance occurrence X (sample data) is observed, then P (X | H) is the chance of a data sample X, which is assumed that the hypothesis H is true (valid).

The probability of X and H are happening simultaneously symbolized by P(X∩H) or P(H∪X). The probability P (X | H) occurs if the events X, occurs when preceded events H, so that its value can be calculated using the equation:

$$P(X|H) = \frac{P(X \cap H)}{P(H)} \quad (1)$$

In the same way, if the event H occurs, preceded by the events of X, then the value of the probability P (H | X) can be calculated by the equation:

$$P(H|X) = \frac{P(H \cap X)}{P(X)} \quad (2)$$

Since P (X∩H) = P (H∩X), it is obtained:

$$\quad (3)$$

$$P(X|H) x P(H) = P(H|X) x P(X)$$

Thus obtained:

$$P(H|X) = \frac{P(X|H) x P(H)}{P(X)} \quad (4)$$

Equation (4) is the one who later became the basis for the Naive Bayes method. In Naive Bayes, because each attribute is assumed to be unrelated from each other (conditionally independent), then the equation can be expressed as follows:

$$P(X|C_i) = \prod_{k=1}^{n} P(X_k|C_i) \quad (5)$$

Based on equation (5), then the class (label) of the data sample X is a class (label) which has:

$$P(X|C_i) * P(C_i) \quad (6)$$

With maximum-value.

## 4. SELF ORGANIZING MAP (SOM)

SOM is a tool that is able to visualize, from which the data are high-dimensional into a low-dimensional, through the reduction process the amount of training data, an increase in the speed of the learning process is done, both for the problem of interpolation and extrapolation is nonlinear and able to perform the compression of the delivery process certain information [22].

SOM contains map units in the form of a grid of two-dimensional (2-D), where each unit *i*, represented by prototype vector $m_i = [m_{i1}, ...., m_{id}]$, where *d* is the dimension of the input vector. Each unit is connected with each other based on the proximity of existing relationships (neighborhood), and trained iteratively, where each training step, the vector *x* is taken at random from the input data sets were then calculated the distance between *x* and all vectors that exist to obtain the Best Matching Unit (BMU), denoted by *b* suppose that the unit map with the closest prototype to *x* (www.mathwork.com). BMU search conducted by the equation:

$$\|x - m_b\| = min\{\|x - m_i\|\} \quad (7)$$

Furthermore, do change the prototype vectors, where the BMU and the topology of adjacent moving closer to the input vector in the input space. Rule changes to the prototype vector of *i-th* unit performed according to the equation:

$$m_i(t+1) = m_i(t) + \alpha(t)h_{bi}(t)[t - m_i(t)] \quad (8)$$

where
t : time
α(t) : the coefficient changes
$h_{bi}(t)$ : the kernel center nearest to the winner unit, where

$$h_{bi}(t) = \exp\left(-\frac{\|r_b - r_i\|^2}{2\sigma^2(t)}\right) \quad (9)$$

Where $r_b$ and $r_i$ is the position of the neuron b and i in the SOM grid. While α(t) and σ(t) decrease monotonically time. In the case of discrete data and the neighborhood kernel is fixed, the error function is determined by the following equation SOM:

$$E = \sum_{i=1}^{N} \sum_{j=1}^{N} h_{bj} \|x_i - m_j\|^2 \quad (10)$$

Where N is the number of training samples, and M denotes the number of units of the map. Neighborhood kernel $h_{bj}$

centered on the unit *b*, which is to the BMU vector $x_i$ and evaluated for the unit to *j*.

## 5. RESEARCH METHODOLOGY

This study began with the collection of data indicators Gross Regional Domestic Product (GDP), both the Provincial level (Banten and Central Java) and district in it. Data collected from the Central Statistic Department (Banten and Central Java) which consists of variables Agriculture, Mining and Quarrying, Manufacturing, Electricity-Gas and Water, Building, Commerce-Hotel and Restaurant, Transport and Communications, Finance-Leasing and services agency, as well as services. The data obtained were then calculated the average percentage growth and the average contribution of the construction according to the districts and provinces.

The data was then entered into the model, where, first, the data will be evaluated using the SOM-NN to produce clusters that will become targets for learning Class performed on the same data when using naive bayes. Furthermore, the learning is done on the same data on Naive Bayess, using Class targets results of the evaluation of learning SOM-NN, then testing the new data to see the results of the classification are formed. In this study, statistical data such as the average value of the growth and development contribution of Central Java Province and District serve as the training data with a total of 135 datasets, while data Banten Province and District totaling 54 data sets, used as a data testing to see the results of the classification of the number of input data.

## 6. RESULT AND DISCUSION

Grouping models are built is a combination of unsupervised techniques (SOM-NN) and supervised (Naive Bayess) to form the final grouping of the data sector GDP. Training data (1) was first evaluated using the SOM-NN (2) becomes a model for classification (3), then, the result of the classification is then used as a new set of data (4) for the evaluation of the target Klass on training data using Naive Bayess.

If the classification result of SOM-NN models has not met, then the process continues training in the model (5) to generate the output of the Class, to be used in the next evaluation stage (6). Evaluation using Naive Bayess used against the same training data (7), thus forming a classification model of Naive Bayess (8) with Class targets, in the form of data sets assessment done by SOM-NN (9). To test the model is formed, then the data testing (10) used the model of Naive Bayess (11) to see the final classification results were generated (12) (see figure 1).

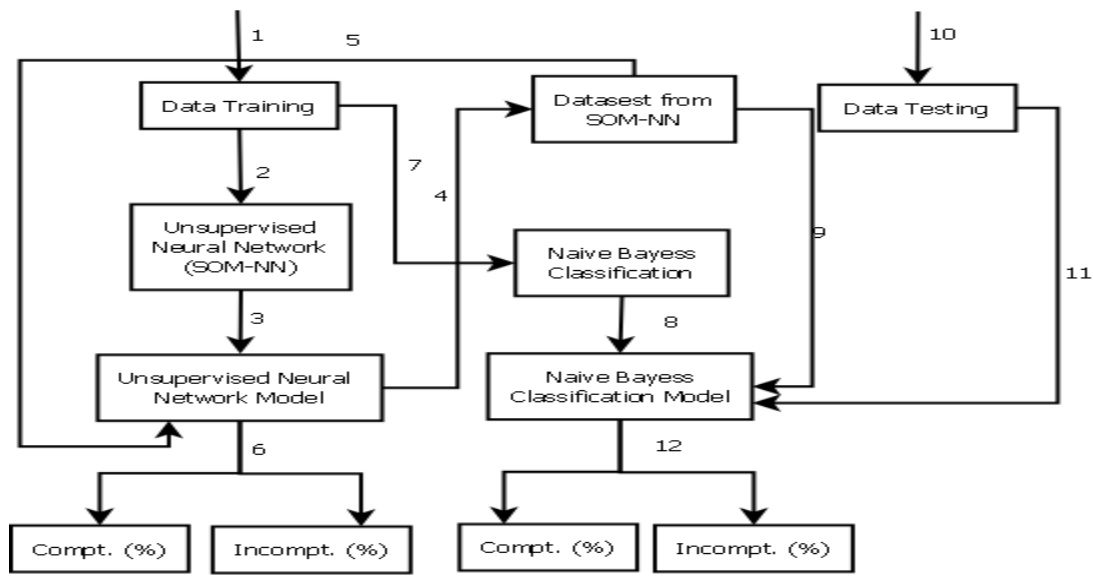

**Fig 1: Proposed Classification Model Architecture**

Simulation data into the model is done using MATLAB applications, by using SOM and Naive Bayess functions provided therein with some additional configuration. Testing the data used is the average data growth and the average contribution of Central Java development with the amount of training data as much as 135 while testing the data used is the average growth rate and the average contribution of the construction of Banten Province with total amount of data testing 54. In testing SOM, the topology used two-dimensional layer with a hexagonal shaped layer of 2x2 so that the number of neurons used 4 pieces with 4 output neurons (see figure 2), as well as the number of epoch is used for the learning process as much as 1000 times.

Learning outcomes using SOM-NN, forming a pattern of data distribution in a hexagonal shape with four Class is formed (see Figure 3) from the maximum loop wherein each Class shows the relationship similarity and disimilarity of training data used.

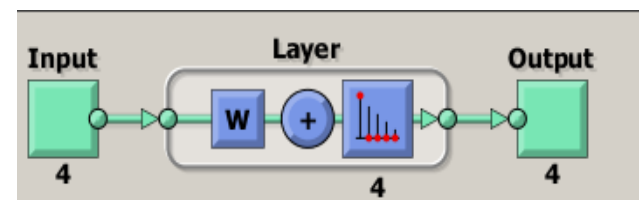

**Fig 2: Self Organizing Map NN Architecture**

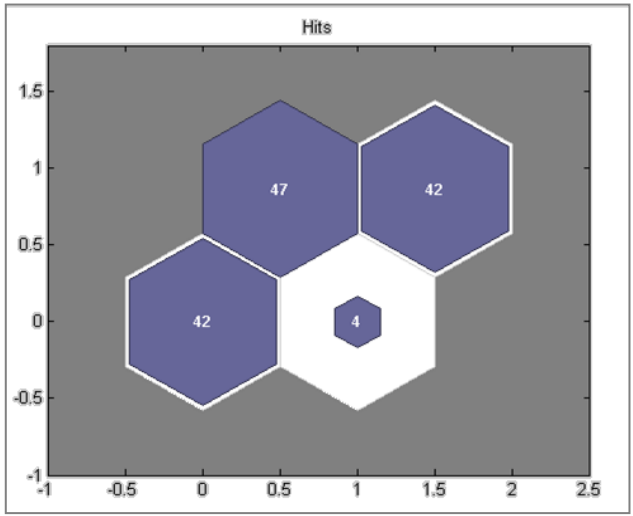

**Fig 3: Topology Generated Class**

Learning outcomes using SOM-NN indicates that the training data is divided into four Class, First Class by the number of members of as many as 42 data sets, the second Class, as many as 4, Class three, as many as 47 and, fourth Class, as many as 42 datasets. This class, which became the new data sets to be used as a target class in learning Naive Bayess.

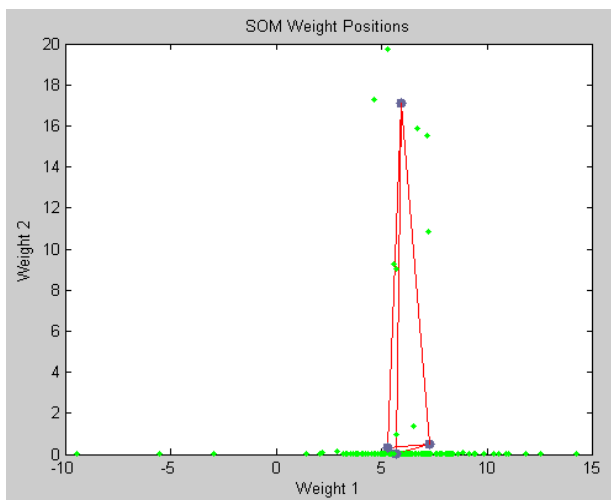

**Fig 3: SOM Weight position**

Figure 4 shows how the SOM to classify the input space to the weight vector of each neuron, as shown green dots, as the input space and the blue dots gray as neurons. While the red line is the liaison between the neuron with other neurons.

Learning is then performed using the method of Naive Bayess with the same training data, with the target Class, the output of SOM-NN learning. In this section, testing of the data directly testing done to see the results of the classification are formed. Testing the data used is the average data growth and average GDP contribution of Development sector data Banten Province as many as 54 data sets. Results of classification using Naive Bayess testing indicates that the data is divided into three four Class Class of the existing target. Outpus Class shows that, on plate formed for testing the data only includes Class first, second and third.

**Table 2. Klassen classification comparison matrix between the proposed models**

| No | Input | | | | Output/Target | |
|----|-------|-----|-----|-----|---------------|---|
|    | V1 | V2 | V3 | V4 | Hasil Klassen | Model Results |
| 1 | 0,871 | 1,56 | 4,31 | 7,27 | 4 | 1 |
| 2 | 6,894 | 0,07 | 6,89 | 0,11 | 3 | 1 |
| 3 | 6,727 | 7,25 | 3,15 | 4,92 | 1 | 3 |
| 4 | 3,056 | 5,28 | 6,36 | 3,66 | 2 | 3 |
| 5 | 7,061 | 0,33 | 955,95 | 279,96 | 4 | 3 |
| 6 | 1,016 | 1,19 | 1,13 | 1,96 | 4 | 1 |
| 7 | 4,839 | 4,93 | 1,04 | 9,16 | 4 | 3 |
| 8 | 8,356 | 2,26 | 7,84 | 3,73 | 3 | 3 |
| 9 | 953,6 | 1,15 | 8,54 | 4,43 | 3 | 3 |
| 10 | 4,279 | 0,16 | 4,31 | 7,27 | 4 | 3 |
| 11 | 0 | 0,00 | 6,89 | 0,11 | 4 | 1 |
| 12 | 2,832 | 4,64 | 3,15 | 4,92 | 4 | 3 |
| 13 | 5,504 | 0,93 | 6,36 | 3,66 | 4 | 1 |
| 14 | 8,809 | 2,04 | 955,95 | 279,96 | 4 | 3 |
| 15 | 9,737 | 3,06 | 1,13 | 1,96 | 2 | 3 |
| 16 | 1,036 | 139,59 | 1,04 | 9,16 | 2 | 2 |
| 17 | 7,997 | 3,70 | 7,84 | 3,73 | 3 | 3 |
| 18 | 8,814 | 2,23 | 8,54 | 4,43 | 3 | 3 |
| 19 | 4,229 | 8,66 | 4,31 | 7,27 | 2 | 3 |
| 20 | 2,666 | 0,02 | 6,89 | 0,11 | 4 | 1 |
| 21 | 7,467 | 4,72 | 3,15 | 4,92 | 3 | 3 |
| 22 | 8,613 | 1,41 | 6,36 | 3,66 | 3 | 3 |
| 23 | 5,805 | 2,24 | 955,95 | 279,96 | 2 | 3 |
| 24 | 8,091 | 254,36 | 1,13 | 1,96 | 2 | 2 |
| 25 | 9,914 | 6,74 | 1,04 | 9,16 | 4 | 3 |
| 26 | 6,608 | 9,42 | 7,84 | 3,73 | 2 | 3 |
| 27 | 7,459 | 2,12 | 8,54 | 4,43 | 2 | 3 |
| 28 | 5,193 | 1,01 | 4,31 | 7,27 | 1 | 3 |
| 29 | 8,127 | 0,10 | 6,89 | 0,11 | 3 | 1 |
| 30 | 5,185 | 5,92 | 3,15 | 4,92 | 1 | 3 |
| 31 | 3,126 | 7,88 | 6,36 | 3,66 | 2 | 3 |
| 32 | 1,002 | 0,79 | 955,95 | 279,96 | 3 | 3 |
| 33 | 11,325 | 9,27 | 1,13 | 1,96 | 3 | 3 |
| 34 | 1,075 | 9,16 | 1,04 | 9,16 | 1 | 2 |
| 35 | 8,212 | 0,32 | 7,84 | 3,73 | 3 | 3 |
| 36 | 8,07 | 3,10 | 8,54 | 4,43 | 4 | 3 |
| 37 | 2,702 | 3,63 | 4,31 | 7,27 | 2 | 3 |
| 38 | 7,718 | 1,26 | 6,89 | 0,11 | 1 | 1 |
| 39 | 3,994 | 8,65 | 3,15 | 4,92 | 3 | 3 |
| 40 | 68,903 | 0,41 | 6,36 | 3,66 | 3 | 3 |
| 41 | 8,081 | 4,60 | 955,95 | 279,96 | 2 | 3 |
| 42 | 6,417 | 2,46 | 1,13 | 1,96 | 2 | 1 |
| 43 | 7,224 | 6,32 | 1,04 | 9,16 | 4 | 3 |
| 44 | 7,464 | 488,35 | 7,84 | 3,73 | 2 | 2 |
| 45 | 6,23 | 1,30 | 8,54 | 4,43 | 2 | 3 |
| 46 | 311,95 | 3,07 | 4,31 | 7,27 | 2 | 3 |
| 47 | 9,354 | 0,13 | 6,89 | 0,11 | 1 | 1 |
| 48 | 3,782 | 1,09 | 3,15 | 4,92 | 3 | 1 |
| 49 | 3,31 | 3,23 | 6,36 | 3,66 | 3 | 1 |
| 50 | 6,336 | 5,08 | 955,95 | 279,96 | 2 | 3 |
| 51 | 5,443 | 2,53 | 1,13 | 1,96 | 2 | 1 |
| 52 | 8,45 | 644,90 | 1,04 | 9,16 | 4 | 2 |
| 53 | 5,965 | 5,55 | 7,84 | 3,73 | 2 | 3 |
| 54 | 5,616 | 1,27 | 8,54 | 4,43 | 2 | 3 |

In Table 2, V1 is the variable for average growth, while V2 shows the average value of the contribution of the construction of the district, V3 and V4 respectively an average growth rate and the average contribution of the development owned by the province. Comparison of cluster results show, there are some results of the classification model dataset is built, showing the same classification number classification typology Klassen, as shown in 8,9,16,17,18,21,22,24,32,33,35 dataset, 38,39,40,44 and 47. Comparison of the results showed similar clusters of 22.63%.

The experimental results further demonstrate that the classification output of the model, determined by the amount of training data and the diversity of data held training data, especially related to the similarity and disimilarity data. The distance between the data with the other data, affect the learning process for the formation of Class, especially when the learning is done using the SOM-NN. The second experiment was tried instead to treat the data, where data sets Banten province, amounting to 54, used as training data to generate the target Class used in Bayess Naive method, while the data of Central Java as much as 135 data sets used as the testing of data, in the model.

The second experiment shows that output SOM-NN learning results formed Class, First Class is dominated by four Class followed by the comparison far enough. While the final classification results of the model, indicating that forms on plate refers to a single Class, namely First Class.

## 7. CONCLUSIONS

Based on the experiments conducted, the proposed model is able to perform the data clustering development results by sector GDP so it can be grouped into classes according to certain predetermined. Learning outcomes using two methods in the proposed model shows the accuracy of the classification results of 98.1%. Nevertheless, the results of this grouping has a level of accuracy with very long intervals when compared with the results of manual analysis using the typology classification Klassen.

The study of the model, showing that the classification is formed can be used as a new approach to determine the level of achievement of the development of a region. However, the accuracy of the classification results of the comparison between the typology Klassen with the model, still needs further review so that the accuracy of the classification results for the better.

The result of the grouping of the model is also influenced by the number and diversity of data sets owned. Treatment inversely to the same data in the second experiment showed that the output Class, the SOM-NN learning shows Class first, third and fourth dominate the results, while in the second experiment, the treatment reversed the data, obtained dominant in Class First Class, so the results of grouping of the proposed model to the training data it refers to the first Class.